# INDIRECT DETECTION OF WIMPS

Marc Kamionkowski[†]

*School of Natural Sciences, Institute for Advanced Study, Princeton, NJ 08540*

## ABSTRACT

I review several proposed techniques for indirect detection of weakly interacting massive particles (WIMPs) in the halo. I focus on distinctive signatures from cosmic-ray positrons, antiprotons, and gamma rays produced by annihilation of WIMPs in the galactic halo.

March 1994

## 1. Introduction

There is almost universal agreement on the existence of dark matter in the Universe[1]. Luminous matter contributes only a fraction, $\Omega_{\rm LUM} \sim 0.01$, of critical density. On the other hand, numerous observations suggest that $\Omega$ is in fact much larger. A number of theoretical arguments, such as inflation and the Dicke-Peebles timing coincidence suggest that the Universe is actually flat, $\Omega = 1$.

Although there is considerable debate on exactly how much dark matter there is, observations of flat galactic rotation curves provide incontrovertible evidence for the existence of dark matter in galactic halos, including our own. In general, rotation curves seem to remain flat as far out from the galactic center as are observed. Therefore, although it remains unclear exactly how much mass is entrained in galactic halos, it seems that the mass density contributed by halos is at least $\Omega_{halo} \gtrsim 0.1$. In other words, the dark matter in spiral galaxies outweighs the luminous matter by at least an order of magnitude. Big-bang nucleosynthesis suggests that there are more baryons than are seen, but it also constrains the mass density in baryons to be $\Omega_b \lesssim 0.1$ (Ref. 2). Therefore, it is plausible that there may be some baryonic dark matter in the form of nonluminous massive compact halo objects (MACHOs) such as neutron stars, brown dwarfs, or black holes, but it is difficult to see how baryons could account for all the halo dark matter.

One of the leading candidates for the dark matter is a weakly-interacting massive particle (WIMP). Suppose that in addition to the known particles of the Standard Model there exists a new, yet undiscovered, stable weakly-interacting massive particle, $X$. It is straightforward to show (see, e.g. Ref. 3) that if such a particle exists, it will have a current cosmological mass density in units of critical density given roughly by $\Omega_X h^2 \simeq \langle \sigma_A v \rangle / (3 \times 10^{-27}\,{\rm cm}^3\,{\rm sec}^{-1})$, where $\langle \sigma_A v \rangle$ is the thermally averaged cross section for annihilation of $X$'s into all lighter particles times relative velocity $v$, and $h$ is the Hubble constant in units of 100 km/sec/Mpc.

One can then ask, what annihilation cross section is required to give $\Omega_X \sim 1$? The answer turns out to be a weak scale cross section, i.e., $\sigma_A \sim \alpha^2/(100\,{\rm GeV})^2$, where $\alpha \sim 0.01$. Virtually all particle physicists will agree that there is new physics beyond the Standard Model, and many (if not most) of the best ideas for new physics introduce the existence of a WIMP. For example, a heavy neutrino associated with an extra generation could be the WIMP, but perhaps the most promising WIMP candidate is the neutralino, a linear combination of the supersymmetric partners of the photon, $Z$ boson, and Higgs bosons[4]. Although there can be significant variety in the detailed properties of the WIMP, generically, the interactions of the WIMP are constrained (by $\Omega_X \sim 1$) to be weak scale, and in most models, the mass of the WIMP varies from about 10 GeV to a few TeV.

A number of direct- and indirect-detection schemes are being pursued in an effort to discover WIMPs in the halo. The first class of experiments are laboratory efforts to detect the recoil energy deposited in a low-background detector when a halo WIMP elastically scatters off a nucleus in the detector[5]. The most promising avenue for indirect detection is observation of energetic neutrinos from WIMP annihilation in the Sun and Earth. WIMPs in the halo which accumulate in the Sun and Earth will annihilate therein and produce energetic neutrinos that can potentially be detected by the many high-energy neutrino telescopes currently in operation or construction. I have reviewed this avenue for detection elsewhere[6], so in this lecture, I will instead focus on several possible cosmic-ray signatures of WIMPs in the galactic halo.

Although the WIMP is stable, two WIMPs can annihilate into ordinary matter such as quarks, leptons, gauge bosons, etc. in the same way they did in the early Universe. If WIMPs exist in the galactic halo, then they will occasionally annihilate, and their annihilation products will produce cosmic rays. The difficulty in inferring the existence of particle dark matter from cosmic rays lies in discrimination between WIMP-induced cosmic rays and those from standard "background" sources. As will be argued below, it is quite plausible that WIMPs may produce distinctive cosmic-ray signatures distinguishable from background. It should also be made clear that propagation of cosmic rays in the Galaxy is quite poorly understood. Due to these astrophysical uncertainties, it is difficult to make reliable predictions for a given particle dark-matter candidate, so negative results from cosmic-ray searches cannot generally be used to constrain dark-matter candidates. On the other hand, if observed, these cosmic-ray signatures could provide a smoking-gun signal for the existence of WIMPs in the halo.

## 2. Cosmic-Ray Antiprotons

The best place to look for a distinctive cosmic-ray signature is where the background is smallest. The majority of cosmic rays are protons, and most of the rest are heavier nuclei. Only a very small fraction are antiprotons. Cosmic-ray antiprotons are produced in standard propagation models by spallation of primary cosmic rays on hydrogen atoms in the interstellar medium (ISM)[7]. The exact flux of antiprotons produced by this mechanism actually varies quite a bit in standard propagation models, and the observational situation is equally cloudy. However, there is one feature of the energy spectrum of such secondary antiprotons that is quite generic to standard cosmic-ray models: It is expected that the flux of antiprotons from primary spallation should fall dramatically at low energies, $E_{\bar{p}} \lesssim$ GeV. This is simply because an antiproton at rest must be produced with a large backward momentum in the center-of-momentum frame. This requires a primary cosmic-ray antiproton with a large energy, and the cosmic-ray spectrum falls steeply with energy.

Annihilation of WIMPs, on the other hand, *can* produce low-energy antiprotons[8]. WIMPs will annihilate into quarks, leptons, gauge bosons, etc. which will then hadronize and produce, among other end products, antiprotons. There is no reason why the flux of such antiprotons should decrease dramatically at energies less than a GeV. Therefore, observation of low-energy cosmic-ray antiprotons would provide evidence for WIMPs in the halo.

Calculation of the antiproton flux from WIMP annihilation is straightforward. One assumes that the WIMPs have an isothermal distribution in the halo with a density suitable for accounting for the rotation curves. The flux is proportional to the annihilation rate in the halo. The energy spectrum of the the antiprotons is determined by the fragmentation functions for producing antiprotons from the various annihilation products, which are obtained from Monte Carlos and from fits to accelerator data. Propagation of the antiprotons through the interstellar medium and solar modulation must also be considered.

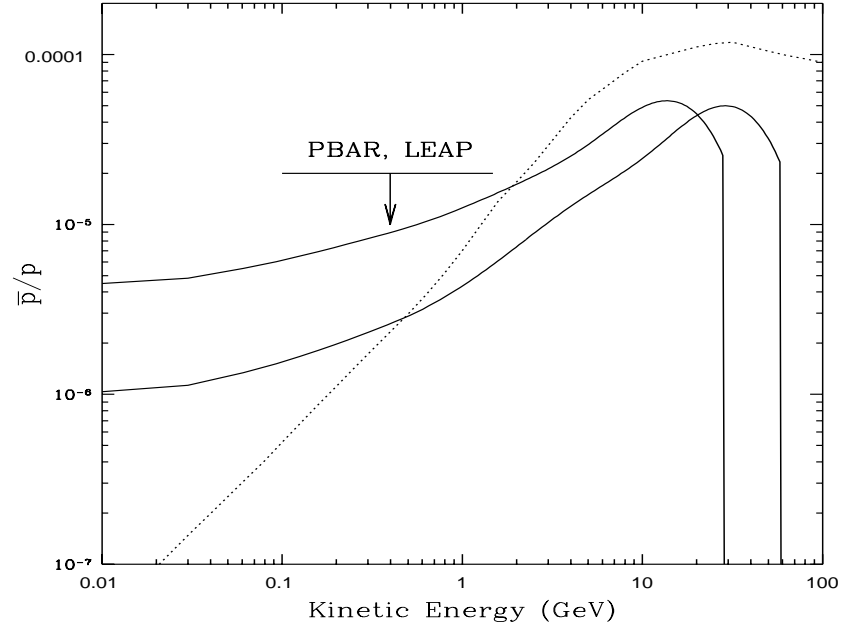

Fig. 1 Observed antiproton/proton ratio as a function of kinetic energy. (From Ref. 8.)

In Fig. 1 are shown the cosmic-ray antiproton spectra expected from models where the dark matter is made up of a $B$-inos of mass 30 GeV (the upper solid curve) or 60 GeV (the lower solid curve)[8]. For simplicity, we chose the WIMP to be a $B$-ino and assumed that the WIMPs contribute closure density, $\Omega_{\tilde{\chi}} h^2 = 0.25$ with $h = 0.5$ to fix the annihilation cross section. We also assumed that WIMPs contribute the entire halo density, and used standard confinement times and solar-modulation models. The dotted curve is the expected background due to spallation in the standard leaky-box model of cosmic-ray propagation. Also shown is the current observational upper limit[9]. As the WIMP mass is increased, the number density in the halo, and therefore the cosmic-ray flux, decrease. As illustrated, observation of low-energy cosmic-ray antiprotons could plausibly provide evidence for the existence of particle dark matter. It should be noted, however, that if the WIMP mass is too large, the antiproton signal would be unobservably small. In addition, even if the WIMP is fairly light, there are considerable astrophysical uncertainties, so it is possible that WIMPs could be the dark matter and still not produce an observable antiproton signal.

## 3. Cosmic-Ray Positrons

There is also a possibility that annihilation of some WIMP candidates will produce a distinctive cosmic-ray positron signature at high energies. Again, there is a "background" of cosmic-ray positrons from spallation of primary cosmic rays off the ISM. Pions produced when primary cosmic rays interact with ISM protons decay to muons which decay to positrons. The flux of positrons, expressed as a fraction of the flux of electrons, decreases slowly with increasing energies.

The showering of WIMP annihilation products will produce positrons in the same way that antiprotons are produced. The energies of the positrons that come from showering of annihilation products will have a broad energy distribution. The background spectrum of positrons expected from standard production mechanisms is quite uncertain, and precise measurements of the positron energy spectrum are quite difficult, so it is unlikely that positrons from WIMP annihilation with a broad energy spectrum could be distinguished from background.

However, in addition to the positrons that come from decays of hadrons, there is also the possibility that WIMPs may annihilate directly into electron-positron pairs thereby producing a "line" source of positrons. Although propagation through the Galaxy would broaden the line somewhat, the observed positron energy spectrum would still have a distinctive peak at an energy equal to the WIMP mass[10]. There are no standard production mechanisms that would produce a positron peak at energies of 10-1000 GeV, so such an observation would be a clear signature of particle dark matter in the halo. It is also interesting to note that some recent measurements of the positron spectrum indicate an increase in the positron fraction at high energies possibly suggestive of WIMP annihilation, although these results are far from conclusive.

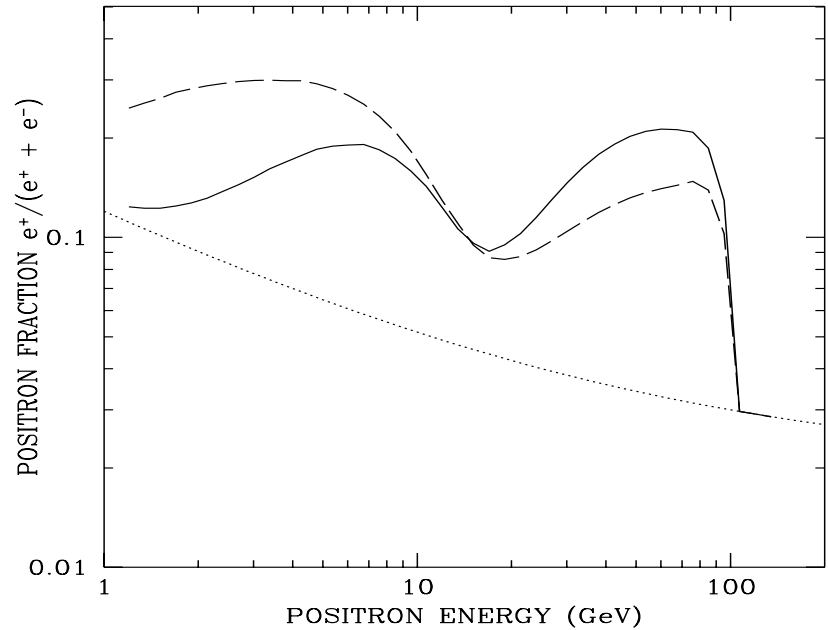

Fig. 2 The differential positron flux divided by the sum of the differential electron and positron fluxes as a function of energy for a neutralino of mass 120 GeV. (From Ref. 11.)

Unfortunately, most of the leading WIMP candidates (e.g. neutralinos) are Majorana particles, and such particles do not decay directly into electron-positron pairs. On the other hand, if the WIMP is heavier than the $W^\pm$ boson, it can in some cases (for example, if the WIMP is a higgsino) annihilate into monochromatic $W^+W^-$ pairs, and the $W^+$ bosons can then decay directly into positrons with a distinctive energy spectrum peaked at roughly half the WIMP mass[11]. In addition, there will be a continuum of lower energy positrons produced by the other decay channels of the gauge bosons.

Fig. 2 shows the differential positron flux as a ratio of the electron-plus-positron flux as a function of energy for a higgsino of mass 120 GeV for two different models of cosmic-ray propagation (the solid and dashed curves). The dotted curve is the expected background. The peak at higher energies is due to direct decays of gauge bosons produced by WIMP annihilation into positrons, and the broader peak at lower energies comes from the other decay channels of the gauge bosons. The dramatic height of the peak in Fig. 2 is the result of some fairly optimistic, yet reasonable astrophysical assumptions. Again, due to the astrophysical uncertainties, nonobservation of such a signal cannot be used to rule out WIMP candidates.

## 4. Cosmic Gamma Rays

Cosmic gamma rays will be produced by annihilation of WIMPs in much the same way that antiprotons and positrons are produced. Showering of the annihilation products will produce gamma rays with a broad energy distribution centered roughly around 1/10th the WIMP mass. Such a signal will in general be difficult to distinguish from background. However, there are two possible signatures of WIMP annihilation in the halo.

The first signature will be a distinctive directional dependence of the gamma-ray flux. In the simplest (and most plausible) models that account for galactic rotation curves, WIMPs populate the halo with a spherically symmetric isothermal distribution. Then, the density $\rho$ of WIMPs as a function of distance $r$ from the galactic center is $\rho(r) = \rho_0(R^2 + a^2)/(r^2 + a^2)$, where $R \simeq 8$ kpc the distance between the solar system and the center of the Galaxy, and $a$ is the scale length of the halo. The ratio $R/a$ varies between roughly 1/3 and 2. Given such a distribution, it is straightforward to calculate the angular dependence of the gamma-ray intensity $I(\psi)$ from WIMP annihilation as a function of $\psi$, the angle between the line of sight and the galactic center. Fig. 3 shows the result for the angular dependence of the gamma-ray flux for three values of the ratio $R/a$. Observation of such a signal would provide evidence for WIMPs in the halo.

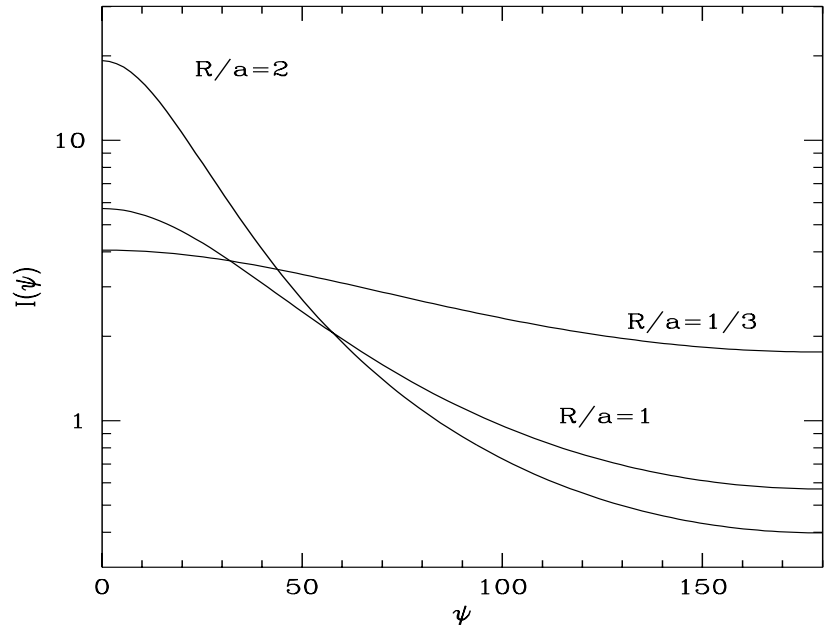

Fig. 3 The intensity of a gamma-ray signal from WIMP annihilation in the halo as a function of the angle between the line of sight and the galactic center. (As in Ref. 12.)

Along similar lines, it has been suggested that there may also be an enhancement in the dark-matter density in the galactic bulge or in the disk and if this dark matter were made of WIMPs, annihilation could lead to a strong gamma-ray signal from the galactic center or the disk[13]; however, it is difficult to see why WIMPs would accumulate at the galactic center or in the disk. Recently, Gondolo has suggested that the Large Magellanic Cloud could be immersed in a halo of dark matter with a central density 10 times that of our own galaxy, and that annihilation of dark matter therein could lead to a gamma-ray intensity from the LMC roughly ten times stronger than that from our own halo[14].

The other, and very distinguishable, signature is a gamma-ray line from direct annihilation of WIMPs into photons. WIMPs, essentially by definition, have no direct coupling to photons. However, by virtue of the fact that the WIMP must have some appreciable coupling to ordinary matter (or else annihilation in the early Universe would be too weak to provide $\Omega_X h^2 \lesssim 1$), it is almost guaranteed that any realistic WIMP will couple to photons through loop diagrams. Therefore, there will always be some small, but finite, cross section for direct annihilation of two WIMPs into gamma rays. Therefore, WIMP annihilation in the halo can produce a gamma-ray signal that is monochromatic at an energy equal to the WIMP mass. There is no easily imaginable astrophysical source that would lead to a gamma-ray line at at an energy between roughly a GeV and a TeV, so discovery of such a line could almost certainly imply the existence of WIMPs in the halo.

The problem with gamma-ray signatures from dark-matter annihilation is that the signals are at best only marginally observable with current detectors even with the most optimistic assumptions. There is, however, hope that heavier WIMPs which couple to the $W^\pm$ boson, such as higgsinos, will annihilate more efficiently into gamma rays[15]. Also, there should be substantial improvements in observational high-energy gamma-ray astronomy in the forthcoming years.

## 5. Conclusions

Of the many proposed dark-matter candidates, the WIMP is perhaps the most promising. The rather suggestive result that a stable particle with weak-scale interactions has a cosmological mass density of order unity has spurred tremendous theoretical and experimental activity in an attempt to detect dark-matter particles. The most reliable detection methods involve terrestrial low-background detectors and searches for energetic neutrinos from WIMP annihilation in the Sun and Earth. However, if WIMPs populate the halo, they will annihilate and produce cosmic rays. Although it will generally be difficult to distinguish such cosmic rays from background, WIMP annihilation may possibly lead to distinctive cosmic-ray signatures. Such signatures are by no means guaranteed even if WIMPs are the dark matter, but in many models it is quite plausible that observations of low-energy antiprotons, high-energy positrons, or gamma rays could provide indirect evidence for the existence of particle dark matter in our halo.

## 6. Acknowledgments

I gratefully acknowledge very productive and enjoyable collaborations with Manuel Drees, Kim Griest, Francis Halzen, Gerard Jungman, Mihoko Nojiri, Tim Stelzer, and Michael Turner. This work was supported by the U.S. Department of Energy under contract DEFG02-90-ER 40542.